# Theoretical construction of thermodynamic relations for a solvent-controlled phase transition to improve the bioavailability of drugs: A case study of indomethacin

K .P. S. de Brito[1,a], T. C. Ramalho[1,b], T. R. Cardoso[2,c] and E .F .F. da Cunha[1,d]

[1] *Department of Chemistry, Federal University of Lavras, PO Box 3037, 37200-000, Lavras- MG, Brazil*

[2] *Department of Physics, Federal University of Lavras, PO Box 3037, 37200-000, Lavras-MG, Brazil*

**Abstract**

The thermodynamic aspects of the polymorphic phase transition from *α*-indomethacin to *γ*-indomethacin are the fundamental key to find the most bioavailable phase of indomethacin. In the present work, varying the temperature and solvent permittivity changes the polymorphic transitions. Hence, the thermodynamic properties such as enthalpy, Gibbs free energy, and entropy of both indomethacin polymorphs are determined in terms of the solvent permittivity as functions of indomethacin's temperature in a vacuum, which are crucially related to the stability, spontaneity, and reversibility of the polymorphic transformation.

**Keywords:** bioavailability, DFT methods (B3LYP/3-21G/PCM), indomethacin, indomethacin polymorphs, phase transition, thermodynamic functions.

**INTRODUCTION**

A drug can crystallize into different polymorphs, which are characterized by their unique spatial arrangement with distinct physicochemical properties, in particular, different solubilities, bioavailabilities and degrees of toxicity [1]. Furthermore, one crystalline arrangement can be transformed into another by the so-called polymorphic phase transition [2]. The understanding on how to induce or suppress polymorphic phase transitions plays a valuable role in the pharmaceutical industry concerning the bioavailability and patentability of the medicine development process [1, 3-5].

[a] kelvyn.brito@dqi.ufla.br
[b] teo@dqi.ufla.br
[c] tati.cardoso@dfi.ufla.br
[d] elaine_cunha@dqi.ufla.br

Notably, indomethacin ($C_{19}H_{16}ClNO_4$) is a very common drug mainly prescribed for its analgesic, anti-inflammatory and antipyretic properties; however, at least seven unique crystalline forms are known. Among them, the α- and γ-indomethacin polymorphs present high stability and allow a crystallographic description, while all of the remaining polymorphs of indomethacin are metastable [6,7]. Here, the thermodynamic properties of indomethacin were investigated in order to find a phase transition driven by the permittivity of the medium [2], provided that a continuous variation in a binary mixture produces a continuous variation in its permittivity value, as expressed in equation (1) [2,8]. The permittivity $\varepsilon_m$ in a binary system composed of two solvents, 1 and 2, whose permittivities are $\varepsilon_1$ and $\varepsilon_2$, respectively, can be expressed as follows:

$$\ln \varepsilon_m = a_1 \ln \varepsilon_1 + a_2 \ln \varepsilon_2 + a_1 a_2 [K_0 + K_1(a_1-a_2) + K(a_1-a_2)^2] \quad (1)$$

where $a_1$ and $a_2$ are the volumetric fractions of solvents 1 and 2 and $K_i$ represent experimental constants [8,9]. The permittivity is independent of the system size and therefore an intensive thermodynamic parameter [10,11].

The importance of this study resides in the fact that, in general, crystals belonging to the same spatial group have similar physical properties [12,13]. The α- and γ-indomethacin polymorphs belong to the space groups $P2_1$ and $P-1$ respectively, which in turn are among the five most common space groups of organic compounds from the Cambridge Crystallographic Data Centre (CCDC). If more than one formula in the asymmetric unit is considered in the space groups (by asymmetric unit, a unit cell composed of more than one chemical compound or molecule, for example, indomethacin and ethanol exist in a single unit cell), then the five most common space groups are the following: $P2_1/c$ (27.8%), $P-1$ (23.5%), $P2_1$ (13.8%), $P1$ (8.5%) and $P2_12_12_1$ (7.8%). On the other hand, if all of the compounds of the CCDC are considered, the following most common space groups are obtained: $P2_1/c$ (36.6%), $P1$ (16.9%), $P2_12_12_1$ (11.0%), $C2/c$ (7.0%) and $P2_1$ (6.4%). Therefore, the generality of the present study is justified by investigating the α- and γ-indomethacin polymorphic phase transition. Accordingly, *ab initio* quantum methods (density functional theory, DFT, methods) could be used to simulate the polymorphic phase transition between α- and γ-indomethacin, which is induced by a range of distinct permittivity values for solvents at a fixed temperature of 298.15 K [6,7]. This range can be precisely controlled by considering the different proportions of solvents in a binary mixture as expressed in equation (1) [8,14]. Hence, the change in the composition of the drug delivery vehicle changes its permittivity, whereas the permittivity and pH values can be related [15,16], and thus, several polymorphs of indomethacin can be experimentally synthesized by altering the pH of the solvent [6], allowing a proper polymorphic transition analysis.

The goal of the present work is to characterize the phase transition of the α- and γ-polymorphs of indomethacin and to establish its thermodynamic properties such as enthalpy, Gibbs free energy and entropy by varying both the temperature and solvent permittivity in DFT computations, which are related to the stability, spontaneity, reversibility and bioavailability of the polymorphs and their transformation.

**METHODOLOGY**

The Gaussian09 program was initially used to determine standard thermodynamic quantities in a vacuum such as the enthalpy (H), Gibbs free energy (G) and entropy (S), which are functions for the α- and γ-indomethacin phases in solvents with different dielectric constant values at a fixed temperature of 298.15 K for validation of the method. The computation was performed by DFT with the B3LYP functional using the 3-21G basis set *via* the polarizable continuum model (PCM) [17-20], using the integral equation formalism variant (IEFPCM) in the default self-consistent reaction field (SCRF) computation. To preserve the number of indomethacin molecules, the original formation of the α-structure inside the solvent consists of the nucleation of six molecules of indomethacin per unit cell, while the nucleation of the γ-structure inside the solvent demands three indomethacin molecules per unit cell [21].

Denoting extensive parameters by $E_1, E_2, ..., E_n$ and intensive parameters by $I_1, I_2, ..., I_n$, a phase transition can be defined by a discontinuity in the derivative of thermodynamic functions (denoted by *P)* in terms of an intensive parameter $I_i$ [22]. In this way, a *n*-order phase transition dependent on $E_i$ is characterized by a discontinuity in the transition point at the (*n-1*)-th derivative of the function $P(E_i(I_i),I_i)$, in which $E_i$ is an extensive thermodynamic parameter that depends on the intensive parameters $I_i$: pressure [23,24], temperature, voltage [25], laser frequency [26], electric field [25], pressure [27,28], permittivity [11] and so forth. The differential form of the thermodynamic function $P(E_i(I_i),I_i)$ can be expressed as $dP = I_1 dE_1 + I_2 dE_2 + ... + I_n dE_n$ [29]. For the present case, the Gibbs free energy is $G = \mu N + D^2/(8\pi\varepsilon)$, and its differential is given straightforwardly by $dG = \mu dN + DdD/4\pi\varepsilon$. Therefore, quantities such as the spontaneity, stability and reversibility of a phase transition can be interpreted by means of $G(E_i(I_i), I_i)$, $H(E_i(I_i), I_i)$ and $S(E_i(I_i), I_i)$, respectively [14,30]. As a particular case, a first-order transition is a discontinuity in the function $P(E_i(I_i), I_i)$ [22]. Therefore, the $\alpha$ to $\gamma$ structural transition can be described as a first-order phase transition in the entropy as a function of the permittivity, as will be shown.

**Table 1:** Characterization of the unit cell of indomethacin polymorphs by the size of its edges (in Å) and angles.

| | | 1-(p-Chlorobenzoyl)-5-methoxy-2-methylindole-3-acetic acid ($C_{14}H_9ClO_3$) | | | | | |
|---|---|---|---|---|---|---|---|
| Polymorph | Space Group | unit cell edge | | | unit cell angle | | |
| | | a | b | c | $\alpha$ | $\beta$ | $\gamma$ |
| α-indomethacin | P2$_1$ (4) | 5.4616(16) | 25.310(9) | 18.152(7) | 90° | 94.38(3)° | 90° |
| γ-indomethacin | P1 (2) | 9.236(5) | 9.620(5) | 10.887(5) | 69.897(5)° | 87.328(5)° | 69.501(5)° |

**RESULTS AND DISCUSSION**

Parameters such as the temperature, solvent permittivity, and pressure in the organic crystal or pH of the medium can significantly modify the stability, solubility and bioavailability of drugs [31,32]. In an effort to improve drug bioavailability knowledge, the solvent permittivity was chosen as the order parameter for thermodynamic analysis. The standard temperature for the polymorphic phase transition between the α and γ phases of indomethacin in the vacuum was reached by the computational method presented in the previous section. The main standard thermodynamic functions, namely, $H(D; \varepsilon, P_0, T_0, N)$, $G(D; \varepsilon, P_0, T_0, N)$ and $S(D; \varepsilon, P_0, T_0, N)$, were computed for six molecules of indomethacin (N=6) in terms of the solvent permittivity $\varepsilon$ at $T_0$=298.15 K and $P_0$=1 atm.

**Theoretical strategy validation: analysis of the enthalpy and Gibbs free energy in a first-order phase transition**

The protocol of the theoretical strategy can be summarized in the correspondence between the simulated temperature for the structural phase transition from the α- to γ-indomethacin polymorphs and the current literature, which are reported in Figure 1.

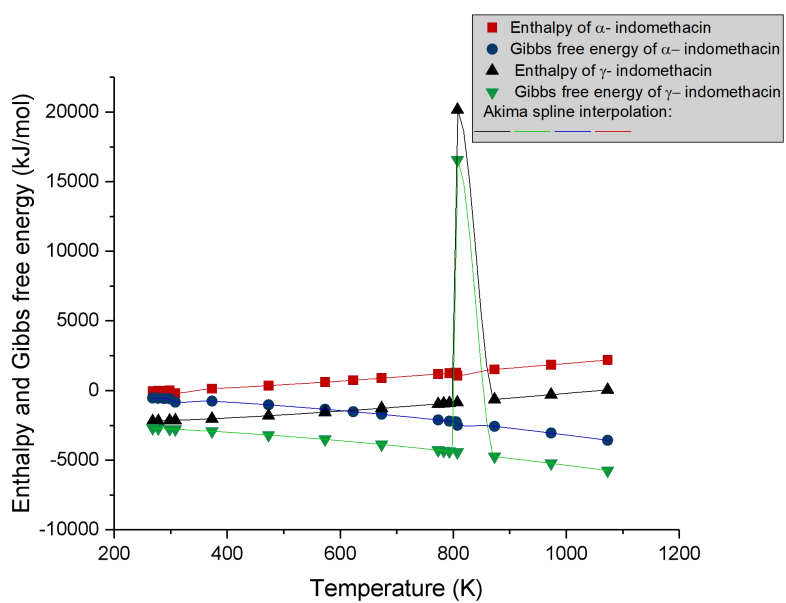

**Figure 1: Enthalpy and Gibbs free energy.** Comparative enthalpy and Gibbs free energy plots for α- and γ-indomethacin.

In Figure 1, the enthalpy for both phases of indomethacin increases with the temperature at a constant rate outside the phase equilibrium, while the Gibbs energy decreases, as expected from the enthalpy expression H=G+TS [29,33].

The critical transition region is highlighted in the direction of the lower Gibbs free energy due to the intersection between both Gibbs free energy curves, characterizing the phase balance, which is corroborated by the current literature, in which the stable phase of γ-indomethacin is achieved at 120 K as well [34,35]. The discontinuity in the Gibbs free energy lies in a temperature range of 793.15 K to 803.15 K due to the presence of a peak for γ-indomethacin.

**Standard thermodynamic parameters as a function of the solvent dielectric constant**

The enthalpy and the Gibbs free energy are plotted in Figure 2 for both indomethacin polymorphs, and their nonlinear behaviors were fitted by the Levenberg Marquardt method, in which all of the coefficients of determination have a correlation given by *R² = 0.999*, in such a way that the quantities $H(D; \varepsilon, P_0, T_0, N)$, $G(D; \varepsilon, P_0, T_0, N)$ and $S(D; \varepsilon, P_0, T_0, N)$, where *S= (H-G)/T,* are all expressed in terms of the permittivity *ε*. Consequently, the theoretical predictions for some transformation properties, such as spontaneity, stability and reversibility, are provided from the standard expressions for Gibbs free energy, enthalpy, and entropy, respectively [30,14].

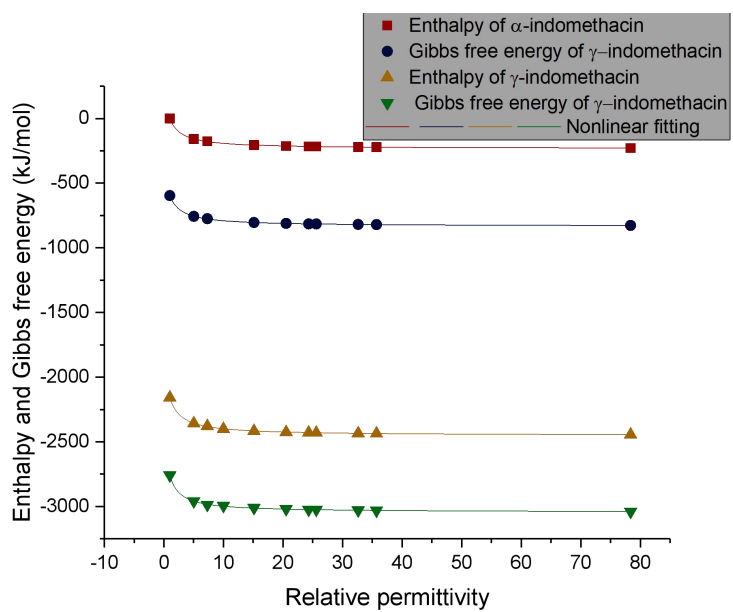

**Figure 2: Enthalpy and Gibbs free energy for different solvents.** Comparative enthalpy and Gibbs free energy plots for α- and γ-indomethacin as a function of the solvent permittivity.

The curves shown in Figure 2 are fitted by the Levenberg Marquardt method $P = \dfrac{b + c\varepsilon}{a + \varepsilon}$, in which *a*, *b* and *c* are constants to be determined. The Gibbs free energy and enthalpy of *γ*-indomethacin are smaller than those of *α*-indomethacin for all permittivity values. Therefore, due to this finding as well as its high degree of symmetry, the *γ*-structure is the most stable and therefore the least bioavailable polymorph of indomethacin [36]. The negative difference between the Gibbs free energies guarantees the spontaneity of the structural transformation from α-indomethacin to γ-indomethacin [9,35]. The parameters obtained from the fitting protocol are expressed in Table 2.

**Table 2**: Parameters from the fitting of the thermodynamic functions

| Model | Rational2 | | | |
|---|---|---|---|---|
| Equation | P = (b + cε)/(a + ε) | | | |
| Plot | Enthalpy of α-indomethacin | Gibbs free energy of α-indomethacin | Enthalpy of γ-indomethacin | Gibbs free energy of γ-indomethacin |
| a | 0.96476 ± 0.03971 | 0.99296 ± 0.0406 | 0.95438 ± 0.04712 | 0.71399 ± 0.07125 |
| b | 233.74839 ± 1.93428 | -355.71274 ± 24.44109 | -1765.40557 ± 101.89131 | -1684.46834 ± 196.6459 |
| c | -233.83429 ± 0.57742 | -833.15349 ± 0.59384 | -2451.09728 ± 0.8954 | -3043.42136 ± 1.38329 |
| Reduced Chi-Sqr | 0.90922 | 0.95559 | 2.19566 | 5.52557 |
| R-Square (COD) | 0.99985 | 0.99985 | 0.99974 | 0.99931 |
| Adj. R-Square | 0.99981 | 0.9998 | 0.99967 | 0.99914 |

Finally, the values for the enthalpy and Gibbs free energy for α- and γ-indomethacin are given by the following equations:

$H_α$ = (-233.83429ε +233.74839)/(ε+0.96476)=-233.834 + 459.342/(0.96476 + ε)      (2)

$H_γ$=-(2451.09728ε +1765.40557)/(ε+0.95438)=-2451.1 + 573.873/(0.95438 + ε)      (3)

$G_α$=-(833.15349ε +355.71274)/(ε+0.99296)=-833.153 + 471.575/(0.99296 + ε)      (4)

$G_γ$=(-3043.42136ε -1684.46834)/(ε+0.71399)=-3043.42 + 488.504/(0.71399 + ε)      (5)

The sign of enthalpy variation (ΔH) between α- and γ-indomethacin characterizes an endothermic (positive) or exothermic (negative) polymorphic transformation. Hence, from equations (2) and (3), the polymorphic transformation from α-indomethacin to γ-indomethacin is observed to be exothermic for all positive values of the permittivity. As an exceptional case, there is no heat exchange if $H_α=H_γ$. The transformation is endothermic if the relative permittivity assumes negative values, which is the case of metamaterials [37-41]. In nanotechnology, negative permittivity values are found in polyaniline–metal oxide (carbon nanotube) composites and in poly(vinylidene fluoride)/Ni [$(CH_2CF_2)_n$/Ni] chain composite films [37]. Mechanisms of formation of negative dielectric properties, which can be applied to drug encapsulation, are also present in the literature [38 - 41].

The polar solvents employed in the energy computation SCRF=PCM are depicted, and the respective permittivity value of each solvent is shown in parentheses: 1-bromooctane (5.0244), aniline (6.8882), pentanal (10.0), 1-pentanol (15.13), 1-propanol (20.524), ethanol (24.852), 1-benzonitrile (25.592), methanol (32.613), acetonitrile (35.688), 1,2-ethanediol (40.245), and water (78.3553).

The spontaneity of the phase transformation can be determined as a function of the Gibbs free energy variation, as described by Figure 3 [29,33].

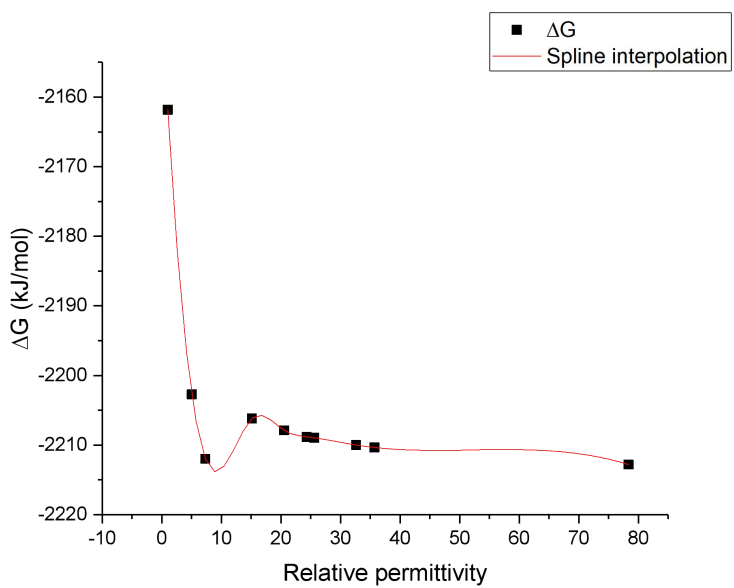

**Figure 3: Gibbs free energy function versus the relative permittivity at the B3LYP/3-21G/PCM level.**

A polymorphic transformation is spontaneous if ΔG < 0 [9], which also gives the direction of transition from α → γ, *i.e.*, establishing that the transformation from α- to γ-indomethacin is spontaneous for all positive values of the permittivity. The transformation reaches equilibrium if ΔG is close to zero, and according to Figure 3, this equilibrium state is reached if the relative permittivity is close to 1 [29,33]. In addition, for some negative values of permittivity, the transformation is reversed, meaning that the transition for the more bioavailable polymorph is favorable [37].

**Influence of the Entropy**

The last thermodynamic function analyzed is the entropy, which provides information about the reversibility of the polymorphic transformation and measures the degree of disorder of the corresponding compound, whose values for both indomethacin forms are computed based on the enthalpy $H(D;\varepsilon,P_0,T_0,N)$ and Gibbs free energy $G(D;\varepsilon,P_0,T_0,N)$ by means of the standard expression $S(D;\varepsilon,P_0,T_0,N) = (H - G)/T$, as outlined in Figure 4. From the second thermodynamic law, the transformation is reversible only if $\Delta S=0$; otherwise, the process is irreversible. Moreover, the entropy of the whole system always increases [29,42]. Hence, the decrease in the entropy of indomethacin results in an increase in the entropy of the solvent counterpart. The change in the sign of $\Delta S$ emerges by means of a discontinuity and occurs at $\varepsilon \approx 11.46$.

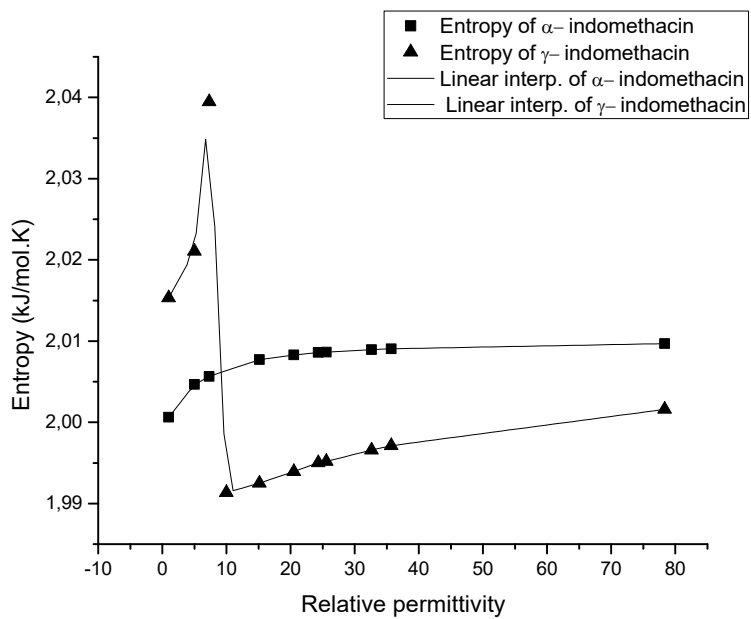

**Figure 4:** Entropy versus relative permittivity at the B3LYP/3-21G/PCM level.

This discontinuity in the entropy characterizes a phase transition driven by the permittivity of the solvent [8]. Moreover, as seen from equation (6), there is also a discontinuity in the first derivative of the entropy:

$$\frac{\Delta S}{\Delta \varepsilon} \approx \frac{\partial S}{\partial \varepsilon} = -\left[\frac{\partial}{\partial \varepsilon}\left(\frac{\partial G}{\partial T}\right)_\varepsilon\right]_T = -\left[\frac{\partial}{\partial T}\left(\frac{\partial G}{\partial \varepsilon}\right)_T\right]_\varepsilon \qquad (6)$$

In addition, the standard relationship between the enthalpy and entropy variations, $\Delta G = \Delta H - T\Delta S$, which can also be visualized in Figure 1, is constant [43], revealing that a phase transition occurs spontaneously for all temperatures in the solvent with a dielectric constant below $\varepsilon \approx 11.5$. On the other hand, if the same conditions were considered for solvents with a dielectric constant above $\varepsilon \approx 11.5$, a very high temperature allows a spontaneous transformation in the opposite direction, from $\gamma$-indomethacin to $\alpha$-indomethacin. In fact, spontaneous transformation can also be achieved by microwave excitation of the hydrogen bonds between molecules, which would be broken, inducing a phase transition [8].

Permittivity is the order parameter that dominates the polymorphic transformations, although pressure should be considered as well. In this case, crystallization occurs at high pressure, and new polymorphs may arise with a higher degree of symmetry and hence less disorder [44]. From another point of view, the same process occurs if the temperature is decreased [45]. Note that $\gamma$-indomethacin is in the space group P-1, thus having a higher degree of symmetry than the $\alpha$-structure and the smallest disorder as a result [44]. Therefore, $\gamma$-indomethacin is produced from $\alpha$-indomethacin if the pressure increases through dry milling [9,10,45]. For this reason, it is speculated that under high-pressure, $\gamma$-indomethacin immersed in solvents whose relative permittivity is greater than 12 transforms into $\alpha$-indomethacin [9,44]. Note that the reversibility aspect of the transformation comes from the difference in the entropy $\Delta S = S_\gamma - S_\alpha$. The entropy of indomethacin decreases in the polymorphic transformation from $\alpha$- to $\gamma$-indomethacin for relative permittivity values greater than 11.46.

**Effect of the solvent permittivity on the solvation Gibbs free energy**

The cycle of the solvation process can be evaluated from Diagram 1.

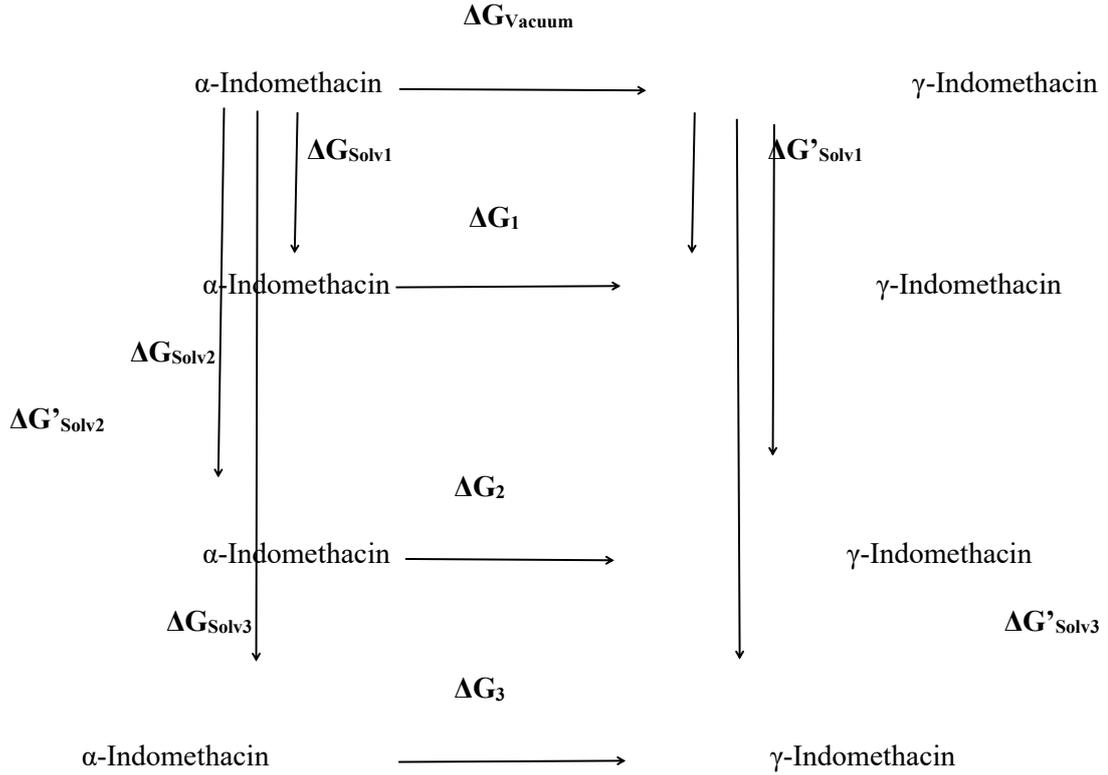

**Diagram 1:** Thermodynamic cycle of solvation Gibbs free energies.

As seen from Diagram 1, the spontaneity of a transformation depends on the difference between the Gibbs free energies of the polymorphs, which can be clarified here with the aid of Figure 7. The solvation Gibbs free energies $\Delta G_{Solv}^{\gamma}$ and $\Delta G_{Solv}^{\alpha}$ are directly related to the difference between the Gibbs free energy of transformation into different solvents and in the vacuum, as established by the following equation [33]:

$$\Delta G = \Delta G_{vac} + (\Delta G_{Solv}^{\gamma} - \Delta G_{Solv}^{\alpha}) \qquad (7)$$

Next, indomethacin treated in different solvents shows diverse values of $\Delta G$ as a result of the medium permittivity, which provides a particular equilibrium constant $K$ for each solvent, as schematized in the following equation [33]:

$$\ln K = -\frac{\Delta G(\varepsilon)}{RT} \quad \Rightarrow \quad K = \frac{[\gamma]}{[\alpha]} = \exp\left(-\frac{\Delta G(\varepsilon)}{RT}\right) \qquad (8)$$

where [α] and [γ] denote the molarity of α- and γ-indomethacin, respectively. Then, the sign of $\Delta G$ determines which polymorph is the most abundant and its proportion. The molarity of γ-indomethacin is greater than that of α-indomethacin if $\Delta G < 0$. Therefore, α-indomethacin can be more abundant than γ-indomethacin in solvents with negative permittivity values. In terms of bioavailability, Figure 3 shows that the molar ratio [α]/[γ] is higher for positive permittivity values, which in turn are close to the vacuum permittivity. The solvation energies can also be assigned as a function of permittivity, as follows.

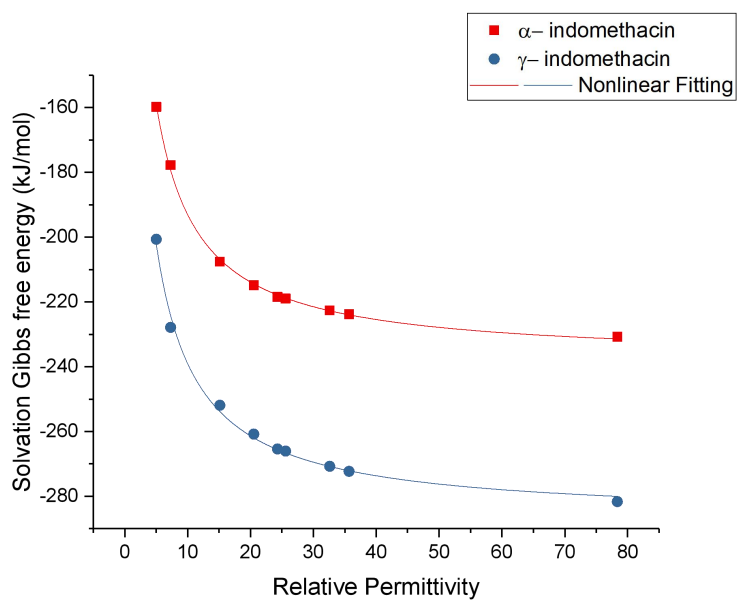

**Figure 5: Solvation Gibbs free energy versus relative permittivity.**

The solvation Gibbs free energy data plotted in Figure 5 provides the parameters from the nonlinear fitting depicted in Table 3.

Table 3: Parameters from the nonlinear fitting of the solvation Gibbs free energy

| Equation | $G_{solv}$ = a+b/(ε-c) | |
|---|---|---|
| Plot | Solvation Gibbs free energy α-indomethacin | Solvation Gibbs free energy γ-indomethacin |
| a | -237.9101 1.003 | -286.89872122.225 |
| b | 518.572881333.83 | 547.467981374.29 |
| c | -1.5525130.3831 | -1.439130.78462 |
| Reduced Chi-Sqr | 0.78542 | 3.94291 |
| R-Square(COD) | 0.99895 | 0.99547 |
| Adj. R-Square | 0.9986 | 0.99396 |

Therefore, Table 3 provides the parameters used to express the Gibbs free energies of solvation:

$$G_{solv,a} = -237.91016 + 518.57288/(\varepsilon+1.5525) \quad (9)$$

$$G_{solv,\gamma} = -286.89872 + 547.46798/(\varepsilon+1.439) \quad (10)$$

The current analysis allows the generalization of the Born equation, as shown in equation (11) [46,33].

$$\Delta_{GSolv} = \beta(-1+1/\varepsilon) \quad (11)$$

where $\beta$ is an experimental constant [31]. The enthalpy, Gibbs free energy and solvation heat can be generalized as follows [33]:

$$P = P_i + c(\varepsilon-\varepsilon_i)^{-1} \quad (12)$$

where $P_i$, $\varepsilon_i$ and $c$ are the experimental constants. The behavior of the thermodynamic function $P$ near a phase transition is characterized by the critical exponent -1. Hence, $P$ can be expanded around the value $\varepsilon=0$, and for very large $\varepsilon$, it is given by the following:

$$P = P_i + c[-(1/\varepsilon_i) - (\varepsilon/\varepsilon_i^2) - (\varepsilon^2/\varepsilon_i^3) + O(\varepsilon^3)] \quad (13a)$$

$$P = P_i + c[(1/\varepsilon) + (\varepsilon_i/\varepsilon^2) + (\varepsilon_i^2/\varepsilon^3) + O(\varepsilon^4)] \quad (13b)$$

Equations (12) and (13) were derived from *ab initio* and first-order computations of a thermodynamic function involving an electric field $D$ medium, in which a term due to the electric energy stored in the dielectric medium of the solvent is incorporated into the thermodynamic structure. The Gibbs free energy is shown by the following:

$$G = U - TS + PV = \mu N + D^2/8\pi\varepsilon \quad (14)$$

Other thermodynamic functions thus come similarly [11].

**CONCLUSIONS**

The present study focuses on the thermodynamic behavior of polymorphic phase transitions between the α and γ phases of indomethacin induced by different polar solvents and in terms of their permittivity. Hence, the enthalpy, Gibbs free energy, entropy and solvation Gibbs free energy were computed at 298.15 K in a nucleation process for various permittivity values of the solvents.

The validation of the DFT method was established by matching the $α$ and $γ$ critical transition temperature at the standard value of 807.45 K, followed by the systematic study of the main direction of the polymorphic transformation, $α→γ$ or $γ→α$, by the variation of the permittivity. The main results provide insights regarding the mechanisms of drug bioavailability and are summarized as follows:

- The enthalpy shows exothermic transformations from α- to γ-indomethacin for each positive value of the permittivity. However, for some negative values of permittivity, characterized by metamaterials [36], the transformation from α to γ is endothermic;

- The Gibbs free energy analysis guarantees the spontaneous transformation from α- to γ-indomethacin and, consequently, a higher concentration of γ-indomethacin for all positive permittivity values. However, for values of ε close to vacuum permittivity, indomethacin presents the highest concentration of α-indomethacin among the positive permittivity values. Moreover, for negative permittivity [37], γ- indomethacin spontaneously transforms into α-indomethacin;

- Due to the discontinuity $(∂S/∂ε)|_{ε-} ≠ (∂S/∂ε)|_{ε+}$, wherein the subscript signs - and + indicate the left and right limits of ε, respectively, there is a first-order phase transition of the entropy at ε = 11.46. For ε < 11.46, the $γ$-indomethacin phase is more disordered; otherwise, α-indomethacin is more disordered when ε > 11.46.

**ACKNOWLEDGEMENTS**

K. P.S. de Brito thanks CNPq for financial support through the project no. 150882/2017-3. T.C. Ramalho thanks CAPES, CNPq and FAPEMIG. T.R. Cardoso acknowledges support from FAPESP through project no. 2016/03921-8. E.F.F. da Cunha thanks CAPES, CNPq and FAPEMIG. The authors are particularly grateful to D.H.S. Leal, J.M. Silla and R. Bufalo for their valuable discussion and advice.